\documentstyle[aps]{revtex}
\font\grbf=cmmib10 
\def\bpi{\hbox{\grbf \char'031}}

\begin{document}
\makeatletter
\title{\bf Quantum tetrahedra and simplicial spin networks}
\author{A.\@ Barbieri}
\address{Dipartimento di Fisica and INFN,\\ 
via Buonarroti 2, I-56100 Pisa, Italy\\
{\rm E-mail:barbier@mailbox.difi.unipi.it}}
\makeatother
\maketitle
\begin{abstract}
\\A new link between tetrahedra and the group $SU(2)$ is pointed out: by 
associating to each face of a tetrahedron an irreducible unitary 
$SU(2)$ representation and by imposing that the faces close, the concept of 
quantum tetrahedron is seen to emerge. The Hilbert space of
the quantum tetrahedron is introduced and it is shown that, due to an 
uncertainty relation, the ``geometry of the tetrahedron'' exists only in the
sense of ``mean geometry''.\\
A kinematical model of quantum gauge theory is also proposed, 
which shares the advantages of the Loop Representation approach in handling 
in a simple way gauge- and diff-invariances at a quantum 
level, but is completely combinatorial. The concept of 
quantum tetrahedron finds a natural application in this model, giving a
possible interpretation of $SU(2)$ spin networks in terms of geometrical 
objects.
\end{abstract}
\section{Introduction}
The links between geometric objects and angular momenta in Quantum 
Mechanics have been observed in many ways since long ago \cite{wig}, 
and the properties of some invariants which can be obtained from $SU(2)$ 
representations have been used by Ponzano and Regge to build a quantum gravity
model in three dimensions \cite{pore}. In the PR model a partition function 
is defined for a given 3-dimensional simplicial complex (it has been proved 
later, by deforming $SU(2)$ to a quantum group \cite{turavi}, that the partition 
function depends only on the topology of the manifold which is triangulated by 
the simplicial complex) by means of the 
following procedure: to each {\em edge} of the complex is associated  a spin 
$j_e$ (that is, an irreducible unitary $SU(2)$ representation,  
characterized only by its dimension $d\equiv 2j+1$). The ``exponential of the 
action'' in such a configuration is a suitable product of the the 
$6j$-symbols associated to the 3-simplices and the partition function is 
obtained by taking the sum of this products over the possible associations of 
spins to the edges.

This construction has been inspired by the following property: ruling out 
the trivial 1-dimensional representations, the $6j$-symbols are zero unless 
the values of the representations' spins can represent lenghts of the edges 
of a tetrahedron (i.e.\@ the triangular inequality is satisfied in each face), 
which is, in a loose sense, the ``building block'' of 
3-dimensional simplicial complexes. The main feature of the PR model is 
that the partition function is defined in a purely algebraic fashion, and in 
the classical limit (i.e.\@ when the spins $j_e$ are large) it resembles  
the Feynman path integral for (Euclidean) General Relativity in 
3-dimensions: one could say that there is a classical geometry (that defined 
by the edges' lengths) obeying a quantum dynamics.

In this paper we highlight another link between tetrahedra and $SU(2)$ by 
associating to each {\em face} of the tetrahedron an irreducible unitary 
$SU(2)$ representation. This construction is, in some sense, dual to that 
of PR and, after the imposition of a ``closure'' condition analogous to the 
``triangular inequality'' condition seen above, it leads to the notion of a 
{\em quantum geometry} by itself. In section \ref{quantet}
we introduce the Hilbert space of the quantum tetrahedron and we discuss some 
of its properties. We shall see that, due to the noncommutativity of some 
operators simultaneously needed in order to define a classical geometry, an
uncertainty relation arises; it is however possible to obtain some 
geometrical informations by taking the mean values of the relevant operators
and we have computed such ``mean geometries'' in some simple cases.

A strict connection exists between quantum tetrahedra and 4-valent 
vertices of $SU(2)$ spin networks. Since their first appearence in the 
pioneering Penrose's work \cite{pen}, aiming to a combinatorial description 
of some kind of quantum geometry, 
spin networks have found applications in many branches of mathematics and
physics; in particular, it is now a firm
result that spin networks embedded into a manifold are ``the skeleton'' of 
the kinematical 
structure of quantum gauge theories with compact gauge groups 
\cite{almmt,baez}. 
When diffeomorphism invariance holds, as is the case in the loop quantization
of (Euclidean, if one wants a compact gauge group) General Relativity using 
Ashtekar connection variables \cite{ash,rosm,rov}, one must factor out all 
the classes of ``equivalent'' spin networks under diffeomorphisms (i.e.\@ one 
must solve the ``momentum constraints'' which arise in the corresponding 
functional quantization), obtaining what are usually 
called $S$-knots. One of the main difficulties with this approach is the 
rather ``knotted'' structure of the set of arbitrary $S$-knots, manifesting 
itself e.g.\@ in the fact that this set is not countable \cite{grott}.

In section \ref{gauge}, we propose a model of 
quantum gauge theory in which the only allowed spin networks are the 
simplest they could be in order to be nontrivial: they are defined on the
(4-valent) dual graphs $\Gamma_M$ of the simplicial complexes $M$ which 
triangulate the manifold. We shall see in this way that both gauge- and 
diff-invariances are obtained, and that the resulting structure is purely 
combinatorial. (We must remark that, while completing this work, we came 
across a similar construction of combinatorial 
quantum gauge theory using simplicial complexes, appeared in recent 
works of other authors \cite{zapa,foti}.) 

We conclude with some speculations about the possible links between our 
model and some ``quantum version'' of Regge Calculus \cite{regge} 
and its possible applications to the problem of Quantum Gravity.
\section{Quantum tetrahedra}\label{quantet}
\subsection{Classical geometry of a tetrahedron}
A tetrahedron can be understood as the convex envelope of four points in 
3-dimensional Euclidean space $E^3$. With reference to figure 1, 
we see that a triad $\vec{e}{}_1$, $\vec{e}{}_2$, $\vec{e}{}_3$
of independent vectors (nine numbers) defines completely the tetrahedron. If 
one is interested only in its properties independently from its space 
orientation (we may consider these properties as defining the ``geometry'' of 
the tetrahedron), the relevant independent parameters, due to the factorization
of the rotation group, become six and can be taken to be 
\begin{equation}
\vec{e}{}_1\cdot\vec{e}{}_1,\ \ 
\vec{e}{}_2\cdot\vec{e}{}_2,\ \
\vec{e}{}_3\cdot\vec{e}{}_3,\ \
\vec{e}{}_1\cdot\vec{e}{}_2,\ \
\vec{e}{}_2\cdot\vec{e}{}_3,\ \
\vec{e}{}_3\cdot\vec{e}{}_1.
\label{indipp}
\end{equation}
Given these six numbers, it is possible to reproduce the original tetrahedron 
apart from its original orientation in space. (The upper bound in this 
hierarchy is given by the twelve cartesian coordinates of the four points 
and the degrees of freedom match: three parameters for the translations group 
plus three for that of rotations plus six of ``geometry''.)

Let us now consider the {\em vectorial areas} of the tetrahedron in figure 
1, given by
\begin{eqnarray}\label{norms}
\vec{n}{}_1 &\equiv &-\vec{e}{}_1\times\vec{e}{}_2,\\
\vec{n}{}_2 &\equiv &-\vec{e}{}_2\times\vec{e}{}_3,\\
\vec{n}{}_3 &\equiv &-\vec{e}{}_3\times\vec{e}{}_1,\\
\vec{n}{}_4 &\equiv &\vec{e}{}_4\times\vec{e}{}_5=-\vec{n}{}_1-\vec{n}{}_2
-\vec{n}{}_3.\label{nfond}
\end{eqnarray}
The last equation, which is simply the ``closure'' condition (in the form
$\int_{\cal S}\vec{n}\,da=0$ it holds for every closed surface 
${\cal S}$),
shows that only three of these vectors, which will be 
called, somewhat improperly, {\em normals}, are independent; it seems 
now natural to see whether the tetrahedron's geometry can be 
reconstructed from the normals rather than from the edges.

The independent	parameters must belong to the set of the invariants 
which can be obtained from $\vec{n}{}_1,\ \vec{n}{}_2,\ \vec{n}{}_3,\ 
\vec{n}{}_4$, that is, their squares (four times the squared areas of the 
faces) and their mutual scalar products (properly normalized, the cosines 
of the dihedral angles associated to the edges). We should add to these 
quantities the triple product $\vec{n}_1\cdot\vec{n}_2\times\vec{n}_3$, 
which is also invariant; this quantity will play a r\^{o}le later.

In all we have ten numbers which, owing to eqn.\@ (\ref{nfond}), are not 
independent; by taking the scalar products of this expression with the normals 
we obtain four independent equations:
\begin{eqnarray}\label{invnfond}
n^2{}_1+n_{12}+n_{13}+n_{14} &=& 0,\\
n^2{}_2+n_{12}+n_{23}+n_{24} &=& 0,\label{n13sost}\\
n^2{}_3+n_{13}+n_{23}+n_{34} &=& 0,\\
n^2{}_4+n_{14}+n_{24}+n_{34} &=& 0,\label{finfond}
\end{eqnarray}
where $n^2{}_i\equiv\vec{n}{}_i\cdot\vec{n}{}_i$ and
$n_{ij}\equiv\vec{n}{}_i\cdot\vec{n}{}_j$ (the square in the first symbol
is part of the notation).

It is easy to verify that independent parameters are the four squared areas 
$n^2{}_i$ and two dihedral angles associated to edges sharing a vertex, 
e.g. $n_{12}$ and $n_{23}$. The relations between couples of angles associated 
to opposite edges are the following:
\begin{eqnarray}
n^2{}_1+n^2{}_2+2n_{12} &=& n^2{}_3+n^2{}_4+2n_{34},\nonumber\\
n^2{}_2+n^2{}_3+2n_{23} &=& n^2{}_1+n^2{}_4+2n_{14},\nonumber\\
n^2{}_1+n^2{}_3+2n_{13} &=& n^2{}_2+n^2{}_4+2n_{24},\nonumber
\end{eqnarray}
and eqn.\@ (\ref{n13sost}), taken into account the last of these expressions, 
gives the following relation between $n_{13}$ and the chosen variables:
\begin{equation}
n_{13}=\frac{1}{2}\left[n^2{}_4-n^2{}_1-n^2{}_2-n^2{}_3\right]-n_{12}-n_{23}.
\label{n13leg}
\end{equation}
By taking the various scalar products between the definitions 
(\ref{norms}-\ref{nfond}), with the parameters (\ref{indipp}) considered
as unknowns, we obtain a system of algebraic equations with just quadratic  
and constant (i.e.\@ functions of $n^2{}_i$ and $n_{ij}$) terms. This system,
provided that some geometrical non-holonomic restrictions such as $n^2{}_i>0$ 
and $|n_{ij}|<(n^2{}_in^2{}_j)^{1/2}$ are satisfied, has two sets of opposite 
real roots; the correct solution is clearly the one with 
$\vec{e}_i\cdot\vec{e}_i>0$, while the other one can be imagined to 
correspond to purely imaginary edges 
($\vec{e}_i\rightarrow{\rm i}\vec{e}_i$).

To summarize, the values of the areas and of two ``non opposite'' dihedral 
angles actually define completely the tetrahedron's geometry.
\subsection{Quantum geometry}
As we have mentioned in the introduction, let us now associate to the four 
faces of the tetrahedron four unitary irreducible representations of $SU(2)$ 
acting on the spaces ${\cal H}_{j_i}$ ($j$ is the spin of the 
representation, while $i$ labels the faces). The ``quantum versions'' of the 
vectorial areas are assumed to be the generators $\vec{\bf J}{}_i$ acting on 
the tensor product
$${\cal H}_{j_1j_2j_3j_4}\equiv\bigotimes_{1=1}^4{\cal H}_{j_i}.$$
In the PR model, the spins associated to the edges have to obey triangular 
inequalities for each face; in our case, we must impose the quantum 
normals to obey the closure condition (\ref{nfond}):
\begin{equation}
\vec{\bf J}{}_1+\vec{\bf J}{}_2+\vec{\bf J}{}_3+\vec{\bf J}{}_4=0,
\label{fund}
\end{equation}
that is, the space ${\cal H}_{j_1j_2j_3j_4}$ must contain a subspace 
${\cal H}^0_{j_1j_2j_3j_4}$ of spherically symmetric vectors.
In such a space the operators 
\begin{equation}
{\bf n}^2{}_i\equiv\vec{\bf J}{}_i\cdot\vec{\bf J}{}_i\ \ \ ,\ \ \ 
{\bf n}_{ij}\equiv\vec{\bf J}{}_i\cdot\vec{\bf J}{}_j
\end{equation} 
are well defined and they obey a set 
of operatorial equations identical to (\ref{invnfond}-\ref{finfond}).
If, in order not to single out particular values of the $j$'s, we consider 
now the orthogonal sum
\begin{equation}
{\cal H}_\tau\equiv\bigoplus_{\{J\}}{\cal H}^0_{\{J\}},
\label{hilquan}
\end{equation} 
where $\{J\}$ runs over the set of {\em ordered} 4-tuples of integers or 
half-integers such that ${\cal H}^0_{\{J\}}$ is nonempty, the 
operators ${\bf n}^2{}_i$ and ${\bf n}_{ij}$ can be defined in 
${\cal H}_\tau$ as hermitean operators, and, since they are blockwise 
diagonal, they continue to obey equations 
(\ref{invnfond}-\ref{finfond}); we thus may say that 
${\bf n}^2{}_i$ and ${\bf n}_{ij}$, acting on  ${\cal H}_\tau$, 
define a {\em quantum tetrahedron}.\\
Condition (\ref{fund}) may be considered to implement in this context the 
fact that the set 
of the faces must form the boundary of the tetrahedron and as such it must 
have no boundary. The topological principle that ``the boundary of a 
boundary is zero'' usually manifests itself in physics via the Stokes 
Theorem 
as an ``inducer'' of automatic consevation laws \cite{MTW}; in our case it 
induces ``automatic'' invariance under rotations imposed onto the 
full tetrahedron: the faces by themselves transform non trivially (since they 
carry a non-zero spin), but the quantum tetrahedron is invariant (the spin 
``created inside it'' must be zero). As we shall see in section \ref{gauge} 
this fact is in intimate connection with gauge invariance.

It is worth noting that, since e.g.\@ $\{{\bf n}^2{}_1,{\bf n}^2{}_2,
{\bf n}^2{}_3,{\bf n}^2{}_4,{\bf n}_{12}\}$ is easily seen to be a complete 
set of commuting operators in ${\cal H}_\tau$, the set of operators which is 
needed in order to determine the geometry is non-commuting. Indeed, if we 
take ${\bf n}_{23}$ as the last parameter we find
\begin{equation}
[{\bf n}_{12},{\bf n}_{23}]=-{\rm i}\vec{\bf J}{}_1\cdot\vec{\bf J}{}_2
\times\vec{\bf J}{}_3\equiv {\rm i}{\bf U},
\label{relcom}
\end{equation}
where the r.h.s., being a scalar quantity, is well defined in 
${\cal H}_\tau$ (the ${\bf n}^2{}_i$ are proportional to the identity in 
each subspace ${\cal H}^0_{\{J\}}$, therefore they commute with all the 
${\bf n}_{ij}$). 
This equation implies the uncertainty relation
\begin{equation}
(\Delta n_{12})(\Delta n_{23})\geq\frac{1}{2}|\langle{\bf U}\rangle|,
\label{indet}
\end{equation} 
where $(\Delta n_{ij})^2\equiv \langle{\bf n}_{ij}{}^2\rangle-
\langle{\bf n}_{ij}\rangle^2$ is the mean square deviation of 
${\bf n}_{ij}$.

\subsubsection*{Permutations and Parity transformation}
Despite its non symmetric definition, the operator ${\bf U}$ is, except for 
the sign, unique: if we take as independent variables e.g. 
${\bf n}_{12}$ and ${\bf n}_{24}$, the commutator is then
$$[{\bf n}_{12},{\bf n}_{24}]=-{\rm i}\vec{\bf J}{}_1\cdot
\vec{\bf J}{}_2\times\vec{\bf J}{}_4=-{\rm i}{\bf U},$$
where the last equality follows from the fundamental equation (\ref{fund}), and
a little reflection shows that $\vec{\bf J}{}_i\cdot\vec{\bf J}{}_j
\times\vec{\bf J}{}_k$ is the same of ${\bf U}$ (recall the definition 
(\ref{relcom}) with the minus sign) if and 
only if $\{ijk\}$ are the first three elements of an odd permutation of 
$\{1234\}$, while it is opposite if the permutation is even. The latter 
property has an interesting connection with the notion of parity.

The description we have given from the beginning of the geometry of 
tetrahedra has skipped the issue of parity: from the {\em numerical values}
alone of the parameters that we have chosen, it is by no means possible 
to distinguish one tetrahedron from its $P$-transformed:
both the parameters (\ref{indipp}) and ${\bf n}^2{}_i,{\bf n}_{ij}$ are 
invariant under the tranformation $\vec{e}_i\rightarrow-\vec{e}_i$. In the 
description in term of the edges this ambiguity can be resolved by the choice 
of a sign for $\vec{e}_1\cdot\vec{e}_2\times\vec{e}_3$ ($=6V$ where $V$ is the 
{\em volume} of the tetrahedron): it is positive if the triad is right-handed,
and negative if it is left handed. Essentially the same thing can be done 
in the description in terms of the normals, but there is a subtlety. 
Classically, the triple product of 
the normals $\vec{n}{}_1,\ \vec{n}{}_2,\ \vec{n}{}_3$, as defined in 
(\ref{norms}-\ref{nfond}), is given by
\begin{equation}
\vec{n}_1\cdot\vec{n}_2\times\vec{n}_3=-(\vec{e}_1\cdot\vec{e}_2\times
\vec{e}_3)^2=-36V^2,
\label{ipovol}
\end{equation}
and it is always negative, whatever 
$\{\vec{e}{}_1,\vec{e}{}_2,\vec{e}{}_3\}$ is right or left 
handed; a closer inspection on the geometrical settings reveals however that 
they point outwards the tetrahedron only in the first case. The reason 
for this is 
that the vector products are defined in terms of the initial basis 
$\{\vec{e}{}_1,\vec{e}{}_2,\vec{e}{}_3\}$ {\em which we have supposed 
right handed} and to mantain the geometrical settings we must change 
sign to all the normals, with the result that the l.h.s.\@ of eqn.\@ 
(\ref{ipovol}) also changes its sign: we may distinguish the original and its 
$P$-transformed tetrahedra by means of the sign we give to 
$\vec{n}_1\cdot\vec{n}_2\times\vec{n}_3$. This prescription, however, 
requires an additional information: an ordering modulo even permutations 
of the normals. Indeed if the normals are non ordered it is not
possible to decide wether a given triplet should be left- or right-handed. 

These considerations about the link between parity and permutations can be 
transposed in the quantum context quite as they stand.
Under the assumption (to be justified later) that ${\bf U}$ is nondegenerate
in each subspace ${\cal H}_{\{J\}}$, an orthonormal basis in 
${\cal H}_\tau$ is also provided by the set $\{|j_1j_2j_3j_4m\rangle\}$ of 
simultaneous eigenvectors of ${\bf n}^2{}_i$ and ${\bf U}$ ($m$ is an 
index that labels the eigenvalues of ${\bf U}$ in 
${\cal H}^0_{j_1j_2j_3j_4}$).  
Since the definition of the space ${\cal H}^0_{j_1j_2j_3j_4}$ is globally 
insensitive to permutations of $\{j_1,j_2,j_3,j_4\}$, it is clear that 
${\bf U}$ has the same eigenvalues in ${\cal H}^0_{\{J\}}$ and in 
${\cal H}^0_{\pi(\{J\})}$, where $\pi$ is a permutation of four elements, and 
that they are symmetric with respect to zero. 
The index $m$ that labels the eigenvalues of ${\bf U}$ in the subspaces 
${\cal H}^0_{\{J\}}$ can thus be chosen to run from to $-(d-1)/2$ to $(d-1)/2$
with integer steps, where $d$ is the dimension of ${\cal H}^0_{\{J\}}$, in 
such a way that the corresponding eigenvalues obey 
$$u(\{J\},m)=-u(\{J\},-m).$$
If we define in ${\cal H}_\tau$ a representation of the group $\Pi^{(4)}$ 
(permutations of four elements) in the following way 
$$\pi\in\Pi^{(4)}\rightarrow\bpi|j_1j_2j_3j_4m\rangle=
|j_{\pi(1)}j_{\pi(2)}j_{\pi(3)}j_{\pi(4)}{\rm sgn}(\pi)m\rangle,$$
where ${\rm sgn}(\pi)=\pm 1$ if $\pi$ is an even (odd) permutation,
we see that the situation is exactly the same found for the classical 
geometry: we may 
define parity on the quantum tetrahedron's Hilbert space in the following way
\begin{equation}
{\bf P}|j_1j_2j_3j_4m\rangle=|j_1j_2j_3j_4-m\rangle,
\end{equation}
so that we have
$${\bf P}{\bf U}{\bf P}=-{\bf U},$$
but without an ordering modulo even permutations of the faces we cannot
decide if ``positive $u$'' means right- or left-handed, since 
\begin{equation}
\bpi^\dagger{\bf U}\bpi={\rm sgn}(\pi){\bf U}.
\label{parit}
\end{equation}
\subsubsection*{Matrix elements of ${\bf U}$}
The spaces ${\cal H}^0_{j_1j_2j_3j_4}$ are all finite-dimensional, and their 
dimension is given by the espression
$$d={\rm Min}\{j_1+j_2,j_3+j_4\}-{\rm Max}\{|j_1-j_2|,|j_3-j_4|\}+1.$$
Since permutations are taken into account by the equivalence 
${\bf U}\sim-{\bf U}$, we may assume $j_1\leq j_2\leq j_3\leq j_4=j$, and 
we have
$$d=j_1+j_2-\frac{1}{2}|j_1+j_4-(j_2+j_3)|+1.$$
The maximum dimensionality $d_{max}=2j+1$ is achieved by the maximally 
symmetric space ${\cal H}^0_{jjjj}$, while the minimum $d_{min}=1$ is 
achieved when any of the $j_j$ is zero or it holds $j_4=j_1+j_2+j_3$ 
(the necessary condition for ${\cal H}^0_{j_1j_2j_3j_4}$ to be non empty is 
$j_4\leq j_1+j_2+j_3$).\\
1-dimensional spaces are peculiar for the following reason:
on those spaces ${\bf U}$, being traceless, is necessarily zero and all the 
parameters needed to define the geometry have a definite value, so that the 
geometry is ``classical''. However, difficulties arise if one tries to 
interpret the results in a classical way; for instance, when $j_1>0$   
the condition $j_4=j_1+j_2+j_3$ corresponds to the 
``quantum sum'' of three parallel vectors, that is, to a ``flat'' (contained 
in a 2-dimensional plane) tetrahedron, while the quantum geometry 
does not lead to this description, since e.g.\@ the sum of the first three 
areas should be equal to the fourth, and one can check that it is not the case.
In the generic case, eqn. (\ref{indet}) implies that the geometry of the
quantum tetrahedron cannot be 
defined exactly, because of ``quantum fluctuations'': we can continue to 
speak about geometry only ``in the mean''.

Let now $j_1,\ j_2,\ j_3,\ j_4$ be ``allowed'' values for the external spins, 
and $\{|j_{12}\rangle\}$ be the orthonormal basis in 
${\cal H}^0_{j_1j_2j_3j_4}$ diagonalizing $(\vec{\bf J}{}_1+
\vec{\bf J}{}_2)^2$ (with the standard choice of the phase factors); the 
matrix element of eqn.\@ (\ref{relcom}) between the 
states $|j_{12}\rangle$ and $|j'_{12}\rangle$ gives:
\begin{equation}
\langle j'_{12}|{\bf U}|j_{12}\rangle=\frac{{\rm i}}{2}
\left[j_{12}(j_{12}+1)-j'_{12}(j'_{12}+1)\right]\langle j'_{12}|
{\bf n}_{23}|j_{12}\rangle,
\label{mtrel1}
\end{equation}
thus the diagonal elements in this basis are zero. 
Concerning the off-diagonal elements, we have found the selection rule 
\begin{equation}
\langle j'_{12}|{\bf U}|j_{12}\rangle\neq 0\ \Rightarrow\  j'_{12}=j_{12}\pm 1,
\label{selrule}
\end{equation} 
but to prove it we need the commutator between ${\bf n}_{12}$ and 
${\bf U}$:
$$[{\bf n}_{12},{\bf U}]=
-{\rm i}\left[({\bf n}^2{}_1+{\bf n}_{12}){\bf n}_{23}-
{\bf n}_{31}({\bf n}^2{}_2+{\bf n}_{12})\right].$$
From the matrix element of this expression between the states 
$|j_{12}\rangle$ and $|j'_{12}\rangle$, with $j_{12}\neq j'_{12}$, 
one obtains
\begin{eqnarray}
\langle j'_{12}|{\bf U}|j_{12}\rangle=
\frac{{\rm i}}{j_{12}(j_{12}+1)-j'_{12}(j'_{12}+1)}
\Big\{&&[j'_{12}(j'_{12}+1)+j_1(j_1+1)-j_2(j_2+1)]\langle 
j'_{12}|{\bf n}_{23}|j_{12}\rangle-\nonumber\\
&&[j_{12}(j_{12}+1)+j_1(j_1+1)-j_2(j_2+1)]\langle 
j'_{12}|{\bf n}_{31}|j_{12}\rangle\Big\},\nonumber
\end{eqnarray} 
and, by substituting in the above expression the operatorial analogue of 
eqn.\@ (\ref{n13leg}), it is found
$$\langle j'_{12}|{\bf U}|j_{12}\rangle={\rm i}\frac{j_{12}(j_{12}+1)+
j'_{12}(j'_{12}+1)}{j_{12}(j_{12}+1)-j'_{12}(j'_{12}+1)}\langle j'_{12}|
{\bf n}{}_{23}|j_{12}\rangle.$$
By comparing this expression with eqn.\@ (\ref{mtrel1}), we see that if
$\langle j'_{12}|{\bf n}{}_{23}|j_{12}\rangle$ is different from zero, then 
it follows that
$$\frac{j_{12}(j_{12}+1)-j'_{12}(j'_{12}+1)}{2}=
\frac{j_{12}(j_{12}+1)+j'_{12}(j'_{12}+1)}
{j_{12}(j_{12}+1)-j'_{12}(j'_{12}+1)},$$
but it is not difficult to show that this equation actually coincides with 
eqn.\@ (\ref{selrule}).\\
${\bf U}$ being self-adjoint, we may restrict our computation to 
$j'_{12}=j_{12}+1$, obtaining
\begin{equation}
\langle j_{12}+1|{\bf U}|j_{12}\rangle=-{\rm i}(j_{12}+1)
\langle j_{12}+1|{\bf n}_{23}|j_{12}\rangle;
\label{mtrell}
\end{equation}
the independent matrix elements 
are then $d-1$, where $d$ is the dimension of ${\cal H}^0_{j_1j_2j_3j_4}$.

The matrix element (\ref{mtrell}) can be expressed in a general form 
by means of the $6j$-symbols;
indeed, the insertion in the r.h.s.\@ of eqn.\@ (\ref{mtrell}) of the 
identity in ${\cal H}^0_{j_1j_2j_3j_4}$ in the form 
$${\bf 1}=\sum_{j_{23}}|j_{23}\rangle\langle j_{23}|$$ 
produces the final result
\begin{equation}
\langle j_{12}+1|{\bf U}|j_{12}\rangle=-\frac{{\rm i}}{2}
(j_{12}+1)\sum_{j_{23}}j_{23}(j_{23}+1)\langle j_{12}+1|j_{23}\rangle
\langle j_{23}|j_{12}\rangle,
\label{mtrel}
\end{equation}
where the scalar products are given by the expression \cite{wig}
$$\langle j_{12}|j_{23}\rangle=(-1)^{j_1+j_2+j_3+j_4}
\sqrt{(2j_{12}+1)(2j_{23}+1)}\left\{\matrix{j_1 & j_2 & j_{12} 
\cr j_3 & j_4 & j_{23}}\right\}.$$
In order to verify if, in practice, the exposed construction produces 
consistent results, we have computed in some simple cases the quantum 
geometries of the quantum tetrahedron via the following procedure:
\begin{itemize}
\item
given the values $j_1,\ j_2,\ j_3,\ j_4$ of the four external spins, we have 
computed the eigenvalues and the (normalized) eigenvectors of the operator
${\bf U}$ in ${\cal H}^0_{j_1j_2j_3j_4}$;
\item
we have then computed, using the mean values of ${\bf n}_{12}$ and 
${\bf n}_{23}$ in each eigenstate of ${\bf U}$ (the choice of these states 
is motivated by reasons of symmetry), the ``mean'' geometry in the form of 
the ``mean'' edges' squared lenghts\footnote
{All the computations have been performed by {\it Mathematica}.}.
The results obtained are shown in the appendix.
\end{itemize}

From the few cases explored we can infer some general properties
of the mean geometries. Firstly, it seems that actually ${\bf U}$ is 
non-degenerate in ${\cal H}^0_{\{J\}}$; secondly, the mean geometries 
corresponding to states with opposite eigenvalues of ${\bf U}$ are the same. 
Furthermore, when there are four or three equal values 
of the external spins, corresponding respectively to equilateral tetrahedra 
and right pyramids with equilateral basis, the mean geometries do not depend
on the eigenvalues, but this ``degeneration'' disappears for less symmetric 
cases.\\
The case (1/2,1/2,1/2,1/2), which in some sense is the first nontrivial one,
is suggestive for the following reason. Since the $\vec{\bf n}{}_i$ must 
have the dimensions of an area, we may write
$$\vec{\bf n}{}_i={\rm l}^2\vec{\bf J}{}_i,$$
where l is a constant with the dimensions of a lenght, and we see that the 
mean length is exactly the unit lenght l.  

In order to have a better understanding of these results it seems necessary 
to have a consistent interpretation of the operator ${\bf U}$. As shown
in 
eqn.\@ (\ref{ipovol}), classically, the triple product 
$\vec{n}_1\cdot\vec{n}_2\times\vec{n}_3$ is (minus) 36 times the squared  
{\it volume} of the tetrahedron. For this reason the operator 
$|{\bf U}|^{1/2}$ could be interpreted as (a multiple of) the quantum 
analogue of the volume. 
It is not obvious, however, that this identification is consistent, 
because (\ref{ipovol}) has been derived from eqns.\@ (\ref{norms}-\ref{nfond}) 
assuming that the edges $\vec{e}_i$ commute.

\subsection{Another dimension?}
The fact that in the highly symmetric cases the mean geometry does not depend
on $m$, suggests that ${\bf U}$ could be interpreted as an object with
higher
dimensional nature, as in Minkowski space an invariant under rotations
may have different 4-dimensional origins. Indeed, in the cases explored, we 
have noticed that the maximum eigenvalue of ${\bf U}$ is always less than
$36\langle V\rangle^2$, where $\langle V\rangle$ is the volume computed using 
the mean geometry.

This property may be used to introduce the concept of a timelike direction. 
(We discuss the timelike case because it is more intriguing, but if one 
wishes the dimension more can be imagined as spacelike; it does not change 
much.)
If we imagine the tetrahedron as imbedded in 3+1 Minkowski 
spacetime, the requirement that all the 
edges are spacelike is equivalent to the requirement that the 4-dimensional
``normal'' of the tetrahedron (i.e.\@ the 1-form $n_a$ dual to the 
trivector $(\vec{e}_1\wedge\vec{e}_2\wedge\vec{e}_3)_{abc}$) lies inside the 
light cone. Assuming a privileged time axis, the immersion properties of our 
spacelike tetrahedron can be
characterized (in an invariant way under spatial rotations) by the ``angle''
formed by its 4-velocity and the time axis, or, simply, by its velocity.
Notice that, due to the condition that $n_a$ lies inside the light cone, 
the velocity is necessarily less than one ($c=1$).\\ 
We than see that, if the inequality
$\langle|{\bf U}|\rangle<36\langle V\rangle^2$ is actually obeyed in all 
cases, it would be natural to interpret $\langle|{\bf U}|\rangle/
36\langle V\rangle^2$ as some power of the ``$\beta$'' of the quantum
tetrahedron with respect to some ``background'' rest frame.
In this light, the symmetry ${\bf U}\sim-{\bf U}$ could be also
interpreted
as the possibility for $n_a$ to lie inside the future or past light cone.

It is clear that if one considers only {\em one} quantum tetrahedron, the 
possibility of defining its velocity with respect to nothing else, can be 
of little use; one the other hand, if we collect togheter some 
quantum tetrahedra joining them by the faces (i.e. we consider suitable 
tensor products of ${\cal H}_\tau$ with itself), it may be then possible 
to define ``spatial slices'' of 4-dimensional ``quantum simplicial 
complexes''. In the next section we will try to formalize in a general 
framework these ideas. 

\section{Simplicial gauge theory}\label{gauge}
The connection between quantum tetrahedra and the spin networks used in the
Loop Representation framework of quantum General Relativity
is fairly obvious: condition (\ref{fund}), 
translated in terms of the representations spins, is nothing but the 
compatibility condition for the spins of the edges adjacent to a 4-valent 
vertex. Furthermore the (eigen)values of the areas of the faces are one 
half those obtained in the Loop Representation framework and 
$|{\bf U}|^{1/2}$ is $2^{1/2}$ times the volume operator of that model, 
restricted to 4-valent vertices (see \cite{rov}). 
This analogy cannot be pushed farther, because in the Loop Representation 
spin networks have vertices whose valence is arbitrary, and for valence 
different from four it does not seem possible to find a geometric 
interpretation in terms perhaps of more complicated polyhedra. 
Anyway, 4-valent vertices seem to play a particular r\^{o}le for the 
following conjunct two reasons: first, the volume operator always acts 
trivially on vertices with valence less than four; second, 
diffeomorphism-equivalence classes of embedded spin networks 
with vertices with valence higher than four are labeled by continuous 
parameters \cite{grott}, thus failing to have a purely combinatorial 
structure.

In this section we shall sketch a model of quantum gauge theory which is a 
kind of 
hybrid between the Loop Representation and lattice gauge theories, and in 
which the only allowed vertices are 4-valent. This model is not complete in an
essential way, since it lacks a dynamics: all the considerations below are 
limited to the kinematical structure, but the framework itself seems to 
suggest how to introduce the concept of a dynamical evolution. We will return 
to this point at the end.  
\subsection{Kinematics}
The whole construction has been inspired by the following simple observation 
(found in \cite{bux}): let $M$ be a 3-dimensional Regge simplicial complex 
with positive metric; then the set
$$M'\equiv M-\Sigma_1(M),$$
where $\Sigma_1(M)$ is the 1-skeleton of $M$, is (isomorphic to) a multiply
connected Riemann manifold, which we will call a {\it Regge manifold}, whose 
fundamental group is isomorphic to the fundamental group of the 1-graph 
$\Gamma_M$ dual to $M$ ($\Gamma_M$ is always 4-valent). This manifold is 
everywhere {\em locally flat} and the information about its ``curvature'' 
is contained into a representation of $\pi(\Gamma_M)$ in $SO(3)$ (or 
$SU(2)$).\\
Turning to gauge theories, all the information about the gauge field is 
contained \cite{barret} into a representation of the group ${\cal L}_0$ 
of base-pointed loops in the gauge group $G$, given by the holonomy of the 
connection (which can be considered as a lagrangian coordinate for a 
3-dimensional theory as well as a canonical coordinate on the 
phase space of a 4-dimensional one; the underlying manifold ${\cal M}$ is 
in both cases 3-dimensional). 
If however the connection is flat, we remain again with the representations
of $\pi_1({\cal M})$ in $G$.

The idea for what should be called a {\em simplicial gauge theory} is to 
``substitute'' the configuration space for a non-flat gauge theory, with the 
set of flat connections which can be defined over the multiply connected 
manifolds obtained by triangulating the ``original'' manifold ${\cal M}$ and 
by removing the 1-skeletons: in this way each configuration is given by a 
couple $(M,h_M)$, where $M$ is a simplicial complex which triangulates 
${\cal M}$ and $h_M$ is a representation of $\pi(\Gamma_M)$ in $G$.
Intuitively one sees that there is no significant ``loss of configurations'', 
as happens when one goes from a non-flat to a flat connection over the 
{\em same} manifold. Furthermore the configuration space has a far more 
manageable form: the part that substitutes the local degrees of freedom,
that is, the structure of the triangulation, is purely combinatorial.
Finally, as any reference to ${\cal M}$ is contained in the properties 
of the simplicial complexes that triangulate it, and is thus of a purely 
topological nature, diffeomorphism invariance (or rather covariance) is, 
despite the ``lattice-like'' formulation, automatically obtained.

It should be pointed out that this construction is simply meaningless if 
interpreted in a classical fashion; in this respect it is not a ``quantized''
model but it has to be ``quantum'' from the beginning. The space of 
states should be taken as the set of the functionals over 
the configuration space; if the gauge group $G$ is compact, and if 
the set of simplicial complexes triangulating a given manifold is 
countable (which is surely the case if ${\cal M}$ is compact), this space 
can be given a natural measure:
$$\langle\phi|\psi\rangle\equiv\sum_M\int_{G^{N(\Gamma_M)}}
\phi^*(M,g_1,\cdots,g_{N(\Gamma_M)})\psi(M,g_1,\cdots,g_{N(\Gamma_M)})
\prod_{i=1}^{N(\Gamma_M)}{\rm d}\mu_i^H,$$
where $N(\Gamma_M)$ is the number of edges of $\Gamma_M$, and ${\rm d}\mu^H$
is the Haar measure for $G$ (the representation $h_M$ is assigned by means 
of the $g_i\in G$ in 1-1 correspondence with the oriented edges of 
$\Gamma_M$).

This measure however is not fully satisfying, because it leaves little 
space to impose additional relations or constraints to handle 
issues like the introduction of a dynamical evolution, the refinements 
of the triangulation and the limit in which one should ``see the continuum''. 
Regarding this problem, we are looking for a generalization of the 
projective techniques used for integration over the gauge groups in 
continuum gauge theories \cite{asht}, using the fundamental groups of 
the manifolds 
$${\cal M}_M\equiv {\cal M}-{\cal T}(\Sigma_1(M))$$
(${\cal T}$ represents a triangulation of ${\cal M}$ ``isomorphic'' to 
$M$) instead of the tame groups. The main obstacle is that the $\Gamma_M$ 
do not form a group in a trivial way.\\ 
Let us now turn to gauge invariance: in perfect analogy with lattice gauge
theories, gauge transformations are defined over each graph of each 
triangulation by assigning an element $U(v)\in G$ to each vertex $v$; for each
edge $e_i$ we have
$$g_i\rightarrow g'_i\equiv U(v^+(e_i))g_iU^{-1}(v^-(e_i)),$$
where $v^+(e_i)$ is the final vertex of $e_i$ and the opposite holds for 
$v^-(e_i)$.

A basis in the space of invariant vectors under gauge transformations is 
provided (see \cite{baez} for the proof) by the 
spin networks defined over the graphs $\Gamma_M$, of which we recall the 
definition in the case of an arbitrary (compact) group $G$; to each edge 
$e$ one assigns a unitary irreducible 
representation $\rho_e$ of $G$, in such a way that for each vertex $v$ the 
following property is satisfied:
let $T(v)$ and $S(v)$ be the set of the edges having $v$ respectively as the 
final and initial vertex; then in the set 
$$\bigotimes_{e\in T(v)}\rho_e\otimes\bigotimes_{e\in S(v)}\rho_e^*,$$
decomposed into a sum of irreducible representations, are contained trivial 
representations. A spin network is identified by such an assignment plus 
a choice of a trivial representation for each vertex.
\subsection{Anything to do with Quantum Gravity?}
When the gauge group is $SU(2)$, a spin network can be interpreted as 
a collection of quantum tetrahedra joined by the faces and 
the concept of a {\it simplicial quantum geometry} for the manifold seems to 
emerge. 
The reason why the ``geometrical content'' of the model has appeared in the 
$SU(2)$ case can be explained by means of the mentioned observation about 
Regge manifolds: gauge transformations are, in this context, simply {\it 
frame rotations} within each tetrahedron, so that there should be no surprise 
that, by handling invariants under these transformations, one gets 
geometrical quantities. Indeed, our belief that the construction exposed 
has, for the $SU(2)$ case, more than something to do with some 
``quantum version'' of Regge Calculus has been recently enforced by some 
observations found in \cite{immir}, where also it is suggested that the natural
``site'' of spin networks in a loop-quantization of Regge Calculus is the 
dual graph rather than the 1-skeleton. 

Another feature of the $SU(2)$ case is that the space of states is a subset 
of the space of states of the Loop Representation of quantum General 
Relativity (each graph $\Gamma_M$ can be imbedded into the manifold by means
of a triangulation), but, as mentioned in the beginning of this section, it 
is the subset that behaves, in the sense specified above, better. The fact 
that the model provides just {\it this} subset is somewhat gratifying and 
gives it some mathematical appeal.

A posteriori, one may ask if it is really necessary to have a classical 
manifold ``in the background''; after all, its only r\^{o}le is to provide 
some constraints on the combinatorial structure of the simplicial complexes 
one must consider. Actually, from the point of wiew of consistency, it does not
seem to make a lot of difference considering all possible (perhaps compact)
simplicial complexes in the space of states, so that the model logically 
decouples from the manifold's concept itself, which, instead, may arise 
as a suitable ``semiclassical limit''. 

Obviously, whether the $SU(2)$ simplicial gauge theory has something to do 
with Quantum Gravity or not, cannot be said before the introduction of some 
kind of dynamics; regarding this latter point, we think there are two 
possible ways that can be pursued.
The first one is the introduction of a ``scalar constraint'' to keep track
of the fact that there is not a background temporal structure; if one tries
to mimic the form of the scalar constraint of General Relativity a problem
arises with the definition of an extrinsic curvature: the gauge group being
$SU(2)$ rather than the Lorentz group, it is not clear how to introduce 
the boosts. One way out could be something analogous to the generalized 
Wick rotation proposed for the Ashtekar formalism \cite{thiem}; we think 
that such a construction could be linked to the property of the operator  
${\bf U}$ which we exposed at the end of the last section.
The second possibility, maybe closest to the ``simplicial spirit'', is the 
definition of transition amplitudes following steps similar to those in 
\cite{smol}, using perhaps the topological moves that generate the 
(classical) evolution in Regge Calculus.    
\bigskip

\acknowledgements
This work is part of my graduation thesis and I wish to thank my supervisor 
Enore Guadagnini for the attention he dedicated to me 
during the last year and Michele Mintchev for his encouragement; I am also 
grateful to Riccardo Giannitrapani for the many helpful discussions we have 
had throughout the years. 

\subsection*{Note added}
A group-theoretical dimensional extension of the PR(TV) model, meanwhile 
appeared in \cite{bacr}, seems to confirm some of the speculations of 
the last unit. It is argued in \cite{craye} that this model have a chance 
to reproduce the euclidean Einstein-Hilbert action in the classical limit.

On the ``minkowskian'' side (that taking into account causality) 
I have to point out that what I called ``second possibility'' above was 
indeed already contained in \cite{foti} (I cannot explain myself my blindness) 
and is used in \cite{smol} to interpret a string worldsheet as a perturbation 
of {\it evolving} ${\cal G}$ simplicial spin networks, where ${\cal G}$ is any 
(quantum) gauge group. It is argued in \cite{smol} that the action of such a 
perturbation with respect to a background which tends (in a still unprecised 
sense) to flat spacetime tends to the spacetime area of the worldsheet.
\appendix
\section{Some mean geometries}
\begin{center}
\begin{tabular}{cccc|l|ll}
$j_1$&$j_2$&$j_3$&$j_4$&\multicolumn{1}{c|}{${\bf U}$ eigenvalues}&
\multicolumn{2}{c}{$l^2_i$}\\
\hline
1/2\  &\  1/2\  &\  1/2\  &\  1/2\  &
$\pm 3^{1/2}/4$& $l^2_1=\cdots=l^2_6=1$&\\

1&1&1&1&$0\ ,\ \pm 3^{1/2}$& $l_1^2=\cdots=l_6^2=2\cdot(2/3)^{1/2}$&\\

3/2&3/2&3/2&3/2&$\pm 3\cdot 3^{1/2}/4\ ,\ \pm 3\cdot35^{1/2}/4$&
$l_1^2=\cdots=l_6^2=5^{1/2}$&\\ 

2&2&2&2&$0\ ,\ \pm 3(11\pm 57^{1/2})^{1/2}/2^{1/2}$& $l_1^2=
\cdots=l_6^2=2\cdot 2^{1/2}$&\\

5/2&5/2&5/2&5/2&$\pm 1155^{1/2}/4\ ,\ $& $l_1^2=
\cdots=l_6^2=(35/3)^{1/2}$&\\
&&&&$\pm(2211\pm 96\cdot481^{1/2})^{1/2}/4$& $l_1^2=
\cdots=l_6^2=(35/3)^{1/2}$&\\
\hline
1&1&1&2&$\pm 3^{1/2}$&$l^2_1=l^2_2=l^2_3=2^{3/2}$&$l^2_4=l^2_5=l^2_6=2^{1/2}$\\

3/2&3/2&3/2&1/2&$\pm 3^{1/2}$&$l^2_1=l^2_2=l^2_3=4$&$l^2_4=l^2_5=l^2_6=1$\\

3/2&3/2&3/2&5/2&$0\ ,\ \pm3^{3/2}$&$l^2_1=l^2_2=l^2_3=4\cdot(5/21)^{1/2}$&
$l^2_4=l^2_5=l^2_6=(35/3)^{1/2}$\\ 

3/2&3/2&3/2&7/2&$\pm3^{5/2}/4$&$l^2_1=l^2_2=l^2_3=3\cdot(3/7)^{1/2}$&
$l^2_4=l^2_5=l^2_6=21^{1/2}$\\

2&2&2&1&$0\ ,\ \pm3^{3/2}$&$l^2_1=l^2_2=l^2_3=5\cdot(2/3)^{1/2}$&
$l^2_4=l^2_5=l^2_6=2\cdot(2/3)^{1/2}$\\

2&2&2&3&$\pm2\cdot3^{1/2}\ ,\ \pm2\cdot30^{1/2}$ &$l^2_1=l^2_2=l^2_3=5/2$&
$l^2_4=l^2_5=l^2_6=4$\\

2&2&2&4&$0\ ,\ \pm6\cdot3^{1/2}$&$l^2_1=l^2_2=l^2_3=19\cdot15^{1/2}/2$&
$l^2_4=l^2_5=l^2_6=7\cdot(3/5)^{1/2}$\\

2&2&2&5&$\pm4\cdot3^{1/2}$&$l^2_1=l^2_2=l^2_3=4\cdot(2/5)^{1/2}$&
$l^2_4=l^2_5=l^2_6=2\cdot10^{1/2}$\\
\hline
1/2&1/2&1&1&$\pm2^{-1/2}$&$l^2_1=l^2_3=l^2_4=l^2_5=23/192^{1/2}$&
$l^2_2=l^{-2}_6=(2/7)^{1/2}$\\

1/2&1/2&3/2&3/2&$\pm15^{1/2}/4$&$l^2_1=l^2_3=l^2_4=l^2_5=11/28^{1/2}$&
$l^2_2=l^{-2}_6=1/7^{1/2}$\\

1/2&1/2&2&2&$\pm(3/2)^{1/2}$&$l^2_1=l^2_3=l^2_4=l^2_5=71/736^{1/2}$&
$l^2_2=l^{-2}_6=(2/23)^{1/2}$\\

1&1&3/2&3/2&0\ ,&$l^2_1=l^2_3=l^2_4=l^2_5=232/23684^{1/2}$&
$l^2_2=l^{-2}_6=(148/143)^{1/2}$\\
&&&&$\pm(13/2)^{1/2}$&$l^2_1=l^2_3=l^2_4=l^2_5=18911/20367464^{1/2}$&
$l^2_2=l^{-2}_6=(2479/2054)^{1/2}$\\

1&1&2&2&0\ ,&$l^2_1=l^2_3=l^2_4=l^2_5=59/440^{1/2}$&
$l^2_2=l^{-2}_6=4\cdot(2/55)^{1/2}$\\
&&&&$\pm11^{1/2}$&$l^2_1=l^2_3=l^2_4=l^2_5=347/8932^{1/2}$&
$l^2_2=l^{-2}_6=(224/319)^{1/2}$\\
\hline
1/2&1&3/2&2&$\pm3/2$&\multicolumn{2}{l}{$\ l_1^2=11/84^{1/2}\ ,\ 
l^2_2=23/1344^{1/2}\ ,\ l^2_3=17\cdot(7/3)^{1/2}/8\ ,$}\\
&&&&&\multicolumn{2}{l}{$\ l^2_4=71/1344^{1/2}\ ,\ l^2_5=16/21^{1/2}\ ,\ 
l^2_6=45\cdot(3/7)^{1/2}/8$}\\
\end{tabular}
\end{center}

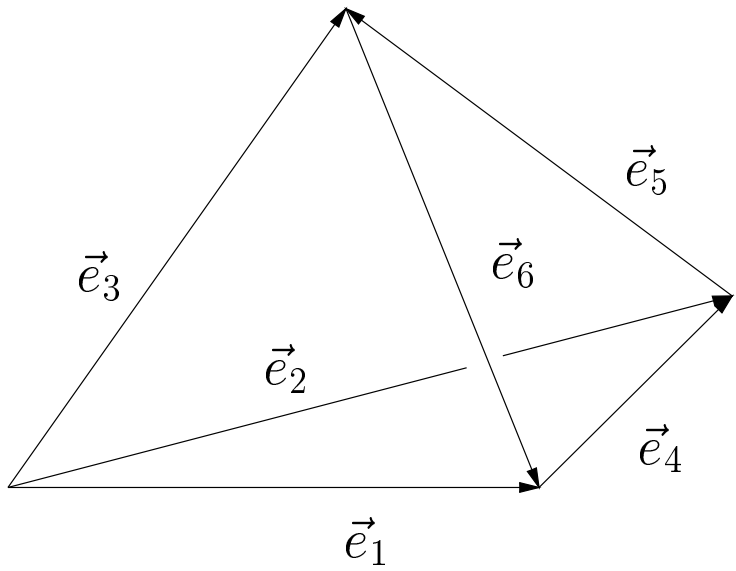
\end{document}